\documentclass[12pt,fleqn]{article}
\usepackage{amsfonts,amssymb}
\title{Gauge Fixing and BFV Quantization}
\author{Alice Rogers\\
Department of Mathematics\\ King's College\\ Strand\\ London WC2R
2LS}
\date{February 1999}
\newcommand{\Bbc}[1]{{\mathbb #1}}
\newcommand{\Bsq}{\blacksquare}
\newcommand{\Seto}{\{}
\newcommand{\Setc}{\}}
\newcommand{\End}{\nonumber\\}
\newcommand{\Comm}{{}[\Omega,\chi]}
\newcommand{\Brst}{\Omega}
\newcommand{\BRST}{{\small BRST}}
\newcommand{\BFV}{{\small BFV}}
\newcommand{\Gff}{\chi}
\newcommand{\Ecomm}{\exp(i \Comm t)}
\newcommand{\Usp}{\Real^{2n}}
\newcommand{\Real}{\Bbc{R}}
\newcommand{\Dpath}{{\cal{D}}}
\newcommand{\Arg}{(p(t),q(t))}
\newcommand{\Pbmat}{(\{ T_a,X^b \} )}
\newcommand{\qdot}{\dot{q}}
\newcommand{\Ham}{H_c}
\newcommand{\Hamt}{\tilde{H}_t}
\newcommand{\Hamg}{H_g}
\newcommand{\Hame}{H_{ext}}
\newcommand{\Hamh}{H_h}
\newcommand{\Expi}[1]{\exp\left[ i #1 t \right]}
\newcommand{\trace}{{\rm Tr}}
\newcommand{\trph}{{\rm Tr}_{{\rm phys}}}
\newcommand{\strace}{{\rm Str}}
\newcommand{\Hil}{{\cal H}}

\newcommand{\Imm}{{\rm Im\,}}
\newcommand{\Ker}{{\rm Ker\,}}
\newcommand{\Cbarstar}{\overline{\gamma}\,{}^*{}}
\newcommand{\astar}{a^*{}}
\newcommand{\bstar}{b^*{}}
\newcommand{\cstar}{\gamma^*{}}
\newcommand{\Cbar}{\overline{\gamma}}
\newcommand{\Cc}{\gamma}
\newcommand{\Rt}[1]{\textstyle{\frac{1}{\surd 2}}( #1)}
\newcommand{\Op}[1]{\hat{#1}}
\newcommand{\kh}{\hat{k}}
\newcommand{\lh}{\hat{l}}

\newcommand{\qh}{\hat{q}}
\newcommand{\ph}{\hat{p}}
\newcommand{\Th}{\hat{t}}
\newcommand{\Theeh}{\hat{\Thee}}
\newcommand{\Etaeh}{\hat{\Etae}}
\newcommand{\Pieh}{\hat{\Pie}}
\newcommand{\Phieh}{\hat{\Phie}}
\newcommand{\Vac}{\Psi_{\scriptstyle{0}}}
\newcommand{\Proof}{{\bf Proof}\ {}}
\newcommand{\Zcoh}{H^0(\Brst)}
\newcommand{\Pcoh}[1]{H^{#1}(\Brst)}
\newcommand{\Endproof}{\hfil $\Bsq$ \break\vskip 0.1cm}
\newcommand{\Pie}{\pi}
\newcommand{\Etae}{\eta}
\newcommand{\Thee}{\theta}
\newcommand{\Phie}{\phi}
\newcommand{\Phin}{\Psi_{mnrs}}
\newcommand{\Phind}{\Psi_{m'n'r's'}}
\newcommand{\half}{\frac12}
\newcommand{\Conj}[1]{\tilde{#1}}
\newcommand{\Frac}[2]{{\displaystyle{\frac{#1}{#2}}}}

\newcommand{\Tone}{{\textstyle{\frac14}}\left((p_1)^2+  (q^1)^2 \right)}
\newcommand{\Ttwo}{{\textstyle{\frac14}}\left((p_2)^2+  (q^2)^2 \right)}
\newcommand{\Xop}{\frac14 \left((p_1)^2 -  (q^1)^2 \right)}
\begin{document}
\newtheorem{Def}{Definition}[section]
\newtheorem{The}{Theorem}[section]
\newtheorem{Lem}{Lemma}[section]
\bibliographystyle{unsrt}
\begin{flushright}
KCL-MTH-99-05
\end{flushright}
\begin{center}
{\LARGE Gauge Fixing and BFV Quantization}\\
\ \\
{\Large Alice Rogers}\\
\ \\
Department of Mathematics\\ King's College\\ Strand\\ London WC2R
2LS
\ \\
February 1999
\end{center}
\begin{abstract}
Nonsingularity conditions are established for the BFV gauge-fixing
ferm\-ion which are sufficient for it to lead to the correct path
integral for a theory with constraints canonically quantized in
the BFV approach. The conditions ensure that anticommutator of
this fermion with the \BRST\ charge regularises the path integral
by regularising the trace over non-physical states in each ghost
sector. The results are applied to the quantization of a system
which has a Gribov problem, using a non-standard form of the
gauge-fixing fermion.
\end{abstract}
%
%
\section{Introduction}
This paper investigates the Fradkin-Vilkovisky theorem, which
establishes the validity of the path-integral expression used in
canonical \BRST\ quantization according to the \BFV\ scheme,
particularly addressing the question of gauge-fixing. The main idea
is to show that the path integral gives a trace over physical
states, with the gauge-fixing term regularising the trace on the
non-physical states by ensuring that these terms cancel out, using
the relation between supertrace and cohomology.

It has long been recognised that in the quantization of gauge
systems some mechanism is required for the cancellation of
redundancy arising from gauge equivalence. The standard method uses
a lagrangian modified by gauge-fixing and Faddeev-Popov ghost
terms, which contribute to the path integral a factor which nicely
divides out the gauge redundancy. The Faddeev-Popov method was
extended by Batalin, Fradkin, Fradkina and Vilkovisky in a series
of papers \cite{fraVil1,BatVil,FraFra,BatFra,FraVil2} to allow
relativistic gauge fixing. (The method developed in these papers
will be referred to as the BFV approach.) Henneaux subsequently
gave an interpretation of the \BFV\  approach in terms of \BRST\
cohomology \cite{Hennea}. There is a detailed account of these
ideas in the book of Henneaux and Teitelboim \cite{HenTei}.

The central idea of these methods is that the original Hamiltonian
$\Ham$ must be adjusted by a term $i\Comm$ where $\Brst$ is the
\BRST\ charge of the theory $\chi$ is the gauge fixing fermion and
$\Comm$ is the super commutator $\Brst\chi+ \chi\Brst$. However
exactly which class of gauge-fixing fermions leads to correct
results has not been entirely clear, as is demonstrated by the
interesting work of Govaerts \cite{Govaer1} and Govaerts and
Troost \cite{GovTro}. The main result of this paper is to
establish the precise conditions which the gauge fixing fermion
$\Gff$ must satisfy if the procedure is to give the correct
result. These conditions (which will be referred to as the {\em
nonsingularity conditions}) are given fully in section
\ref{FVTsec}, together with a proof of the main result. Two
examples are given of the use of this theorem; one is the simple
example of a system with translational invariance in one
direction, the corresponding constraint being that the component
of momentum in this direction is zero. In this case well-known
results are recovered. The second is an example where a Gribov
ambiguity might be expected to prevent application of any
technique involving gauge-fixing; it is shown that in fact a gauge
fixing fermion can be chosen so that the the \BFV\ formalism is
valid.

The proof of the main result depends on the
fact (pointed out by Schwarz \cite{Schwar}) that the supertrace (or,
in the terminology of Schwarz' paper, the Lefschetz trace)of a
\BRST-invariant  operator is equal to the alternating sum of the
traces  of this operator on cohomology classes. This is precisely
the trace over physical states if the only non-trivial \BRST\
cohomology is at ghost number zero.
Then, provided it can be shown that the path integral expression
for the vacuum generating function does give the supertrace of the
evolution operator, the theorem is established. However, since even
in the quantum-mechanical setting taking a trace involves infinite
sums, care must be taken that operators have well-defined traces.
It will be shown that the operator $\Expi{\Comm}$, which is equal
to the identity operator on the space of physical states, regularises
the trace on the non-physical states in such a way that
contributions from these states cancel out.

The paper begins with a brief description of the quantization
method set up by \BFV\  and extended by Henneaux. A simple but
important lemma is then proved, which establishes that the first
nonsingularity condition ensures that the only
states with non-trivial \BRST\  cohomology have ghost number zero.
A study of the eigenstates and eigenvalues of $\Ecomm$ is then
used to show (by arguments similar to those of McKean and Singer
\cite {McKSin}) that the path integral defined by the extended
Hamiltonian $\Hamt + i\Comm$ gives the correct generating
functional. (Here $\Hamt$ is a \BRST-invariant extension of the
original Hamiltonian $\Ham$.) The main result is then applied to
two specific systems; the second of these has a Gribov problem
\cite{Gribov} which makes standard gauge-fixing methods unworkable
but can be handled by the methods of this paper. The difficulties of
gauge-fixing in canonical quantization when there is a Gribov problem
have been analysed by McMullan \cite{McMull}.

\section{BFV Quantization}\label{BFVsec}
The starting point of the \BFV\  approach is the unconstrained
$2n$-dimensional phase space $\Usp$, (with local coordinates
$(p_i,q^i)$, where $i=1,\dots, n$) together with a set of $m$ first
class constraints $T_a(p,q)=0, a=1,\dots,m$ (with $m<n$) and first
class Hamiltonian $\Ham(p,q)$. The true (reduced) phase space of
this system is then the space $B=C/G$ where $C$ is the submanifold
of $\Usp$ on which the constraints hold and $G$ denotes the group
generated by the constraints (which acts naturally on $C$), with
symplectic structure given by the Dirac bracket.
(Finite-dimensional language is used here, but in principle the
ideas can be extended, at least formally, to field theory.) A set
of gauge-fixing functions $X^a, a=1,\dots,m$ are also introduced;
the nature of these functions is a central point of the discussion
of this paper.

Now, as is shown in \cite{Faddee}, the generating functional is
given by the Faddeev formula
 \begin{eqnarray}
   Z &=& \int \Dpath p   \Dpath q   \Bigg[ \prod_t
 \left(\prod_{a=1}^m \left[ \delta \left(T_a\Arg \right) \delta \left (X^a\Arg \right)  \right] \right)\End
 && \times \det \left(\{ T_a\Arg,X^b\Arg \}\right) \End
 && \times \exp \left(  i\int_0^t p_a(t) \qdot^a(t)   - \Ham(p(t)  ,q(t)) dt  \right)\Bigg],
 \label{FFeq}\end{eqnarray}
where the integration  is over
paths $p(t), q(t)$ in the unconstrained phase space which begin
and end at the same point. (Standard quantization of the unconstrained phase space in
the Schr\"odinger picture is used.) In establishing this
expression, further assumptions have to be made about the
gauge-fixing functions $X^a$, including the requirement that the matrix $\Pbmat$
(where $\{ \hspace{7pt}, \hspace{7pt} \}$ denotes the Poisson
bracket) must satisfy the condition
 \begin{equation}
  \det \Pbmat \not= 0
  \label{PBeq}
 \end{equation}
at all points $(p,q)$  in $\Usp$. When it is not possible to find
a set of gauge-fixing functions satisfying these conditions there
is a  Gribov problem \cite{Gribov}, and it is  not possible to
define a unique representative of each class in the  reduced phase
space $C/B$ by picking out the zeroes of a set of gauge-fixing
functions. (An example where this occurs is considered in section
\ref{GRIsec}.)

The key idea in the \BFV\  approach is to extend the phase space by
including Lagrange multipliers $l^a, a=1,\dots,m$ for the
constraints together with their canonically conjugate momenta
$k_a$, a set of $m$ ghosts, $\Etae^a$ together with their conjugate
momenta $\Pie_a$ and a set of $m$ antighosts and corresponding
momenta $\Phie_a$ and $\Thee^a$. An extended Hamiltonian $\Hame
= \Ham +\Hamh +\Hamg$ is then defined, with $\Hamg$ taking a form
prescribed by the constraints, their commutators, and the
gauge-fixing functions, while $\Hamh$ (which is often zero) depends
on the commutator of $\Ham$ with the constraints. The extended
Hamiltonian is quite complicated, but it is shown in
\cite{FraVil2}, by clever manipulation of path integrals, both that
the corresponding generating functional
 \begin{eqnarray}\label{FVSSeq}
  Z &=& \int \Dpath p   \Dpath q   \Dpath k   \Dpath l   \Dpath
    \Pie   \Dpath \Etae   \Dpath \Phie   \Dpath \Thee   \End
  &&  \times \Bigg[ \exp \bigg(i \int_0^t p(t)  \qdot (t)  + k (t)
    \dot{l}(t) +  \Pie(t) \dot{\Etae}(t)   +   {\Phie}(t)    \dot{\Thee}(t)   \quad  \End
    &&\qquad \qquad  -  \Hame(p(t)  ,q (t) ,k (t) , l  (t), \Pie (t) ,
   \Etae(t), \Phie(t) , \Thee(t)  ) dt\bigg) \Bigg] \End
  &&
 \end{eqnarray}
is independent of the choice of gauge-fixing functions and that it
is equal to the Faddeev formula (\ref{FFeq}) for the generating
functional for the original Hamiltonian on the reduced phase space.
This result is known as the {\em Fradkin-Vilkovisky Theorem}.
\section{BRST Cohomology and the Fradkin-\break Vilkovisky
theorem}\label{FVTsec}
A significant insight of Henneaux \cite{Hennea} was an
interpretation of the
\BFV\ Hamiltonian $\Hame = \Ham + \Hamh + \Hamg$ in terms of the
cohomology of the \BRST\  operator $\Brst$ corresponding to the
$2(m+n)$-{}dimensional phase space (with typical point $(p,q,l,k)$)
subject to the constraints $T_a(p,q)=0,\, k_a=0$. Henneaux showed
that the term $\Hamh$ (which contains ghosts)
 led to a modified Hamiltonian
 $\Hamt = \Ham + \Hamh$ satisfying $[\Hamt, \Brst]=0$,
 while the term $\Hamg$ could be expressed as $i\Comm$ with a ``gauge-fixing fermion''
$\Gff$ of ghost number $-1$ constructed from the gauge-fixing
functions $X^a, a= 1, \dots, m$ in a prescribed way. Henneaux also
demonstrated the correspondence between observables on the reduced
phase space and operators which commute with $\Brst$, and the
related correspondence between states for the reduced system and
$\Brst$-cohomology classes.

Henneaux proves the  Fradkin-Vilkovisky theorem by showing that
the path integral (\ref{FVSSeq}) is invariant under infinitesimal
change of gauge-fixing fermi-on. However this is not a full proof,
first, because (as observed by Govaerts \cite{Govaer1}) the space
of orbits of the gauge group may not be connected, and second
because it must be shown that the space of possible gauge-fixing
fermions includes one which gives the correct generating
functional.  (This is achieved in the longer proof in
\cite{FraVil2} by showing that the path integral (\ref{FVSSeq})
reduces to the Faddeev formula (\ref{FFeq}) after integrating out
the fermions and the Lagrange multiplier variables while rescaling
the gauge-fixing functions.)

Now the Faddeev formula (\ref{FFeq}) gives a useful expression for the
reduced phase space generating functional, but involves gauge-fixing functions of a
kind which may not exist for a general system.  However the
existence of the reduced phase space is quite independent of the
existence of such gauge-fixing functions. Indeed, as was shown
quite generally by Kostant and Sternberg \cite{KosSte}, the space
of observables on the reduced phase space is isomorphic to the zero
$\Brst$ cohomology on the extended phase space, so that the space
of physical states can be identified with the zero $\Brst$
cohomology group of states for the extended phase space. This
suggests that it might be possible to prove the Fradkin-Vilkovisky
theorem without reference to the Faddeev formula, and hence without
recourse to gauge-fixing functions satisfying (\ref{PBeq}), and it
is just such a programme which will be carried out here, showing
that there is a connection between the supertrace of an observable
and the \BRST\  cohomology of the states which leads to the
Fradkin-Vilkovisky   theorem, with precise criteria which the
gauge-fixing fermion $\Gff$ must satisfy.

In the \BRST\  cohomology approach to gauge quantization in the
Schr\"od\-ing\-er picture, the full space of states $\Hil$ consists
of functions $f(q,l,\Etae,\Phie)$ of the configuration space
variables. (Precisely which functions are included in $\Hil$
depends on the system under consideration, as will be seen below;
since the inner product on states is not positive definite the
usual $L^2$ prescription will not be valid.) The ghost number of a
state is the degree in the ghost variables $\Etae$ less the degree
in the antighost variables $\Phie$, and physical states are
identified as elements of $\Zcoh$, (where for $i=-m, \dots, m$,
$H^i(\Brst)$ denotes the space of cohomology classes of $\Brst$ at
ghost number $i$). Observables are operators on $\Hil$ which
commute with $\Brst$, and thus have a well defined action on the
cohomology classes $H^i(\Brst)$.

A new proof of the Fradkin-Vilkovisky theorem, with explicit
conditions which the gauge-fixing fermion must satisfy, will now be
given. The {\em nonsingularity conditions}  are that
 \begin{enumerate}
  \renewcommand{\labelenumi}{(\roman{enumi})}
  \item  the only states which are zero eigenstates of the
operator $\Comm$  are states of ghost number zero which are not
$\Omega$-exact;

  \item on each ghost and $\Hamt$ sector  the real part of the eigenvalues
$\lambda_n$ of $i\Comm$ tend to infinity with $n$;

  \item the  Hamiltonian $\Hamt$ must have finite trace on the
  space of zeroes of $\Comm$,  on which it acts modulo $\Omega$.
 \end{enumerate}
As will be seen, the purpose of the first condition is to ensure
that $\Brst$ only has cohomology at ghost-number zero, the second
condition ensures that $i\Comm$ regularises the traces of
non-physical states involved in the path integral and the third
ensures that the system does possess a generating functional.

The first stage of  the proof of the Fradkin-Vilkovisky theorem
given in this section is to show (by two lemmas) that the
supertrace of $\Expi{H}$ gives the trace of this operator over
physical states, that is, the desired generating functional.

For a given operator $A$ on $\Hil$, the supertrace is defined by
 \begin{equation}
  \strace A = \trace \left( (-1)^g\,A \right)
 \label{strtr}\end{equation}
where $g$ denotes ghost number, so that $(-1)^g f=f$ when $f$ is a
state of even ghost number and $(-1)^g f= -f$ when $f$ is a state
of odd ghost number. (We assume for the time being that the various
infinite sums here are well defined; it will be seen below that
this is ensured by the nonsingularity conditions.) Following
Henneaux and Teitelboim (and the earlier work of Schwarz
\cite{Schwar}), it is shown that taking the supertrace of an
observable is equivalent to taking the alternate sum of the traces
over the cohomology classes $H^i(\Brst)$.
 \begin{Lem}\label{STlem}
Suppose that $A$ is an observable on $\Hil$, so that $[A,\Brst] =
0$. Then, assuming that the necessary traces exist,
 \begin{equation}
  \strace A = \sum_{i=-m}^{m} (-1)^i \trace_{H^i(\Brst)} A.
 \end{equation}
 \end{Lem}
\Proof
We make the following decomposition of the space of states $\Hil$:
 \begin{equation}
  \Hil =  G \oplus F  \oplus E
 \end{equation}
where $G = \Imm \Brst$ and $G \oplus E = \Ker \Brst$. Also, for
$i=-m, \dots, m$, let $E_i, F_i$ and $G_i$ denote the subspaces of
$E,F$ and $G$ respectively with ghost number $i$.  Then $G_{i+1} =
\Brst F_i$ for $i=-m, \dots, m-1$, while $G_{-m}$ and $F_{m}$ are
empty since the operator $\Brst$ raises ghost number by one and
annihilates all states of top ghost number.

Now if $f$ is an eigenstate of $A$ with eigenvalue $\lambda$, then
$\Brst f$ is either zero or also an eigenstate of $A$ with the same
eigenvalue; thus the map $\Brst: F_{i} \to G_{i+1}$ is an
isomorphism, so that
 \begin{equation}
 \trace_{F_i} A = \trace_{G_{i+1}} A,
 \end{equation}
which means that the contributions to the supertrace from $F_i$
and $G_{i+1}$ cancel giving
 \begin{equation}
  \strace A = \sum_{i=-m}^{m} (-1)^i\trace_{E_i} A
 \end{equation}
as required. \Endproof
It is next shown that, if $\Gff$ can be chosen so that $\Comm$ has
a certain property, then $\Brst$ only has cohomology at ghost
number zero.
 \begin{Lem}\label{COlem}
If $\Gff$ can be chosen so that the operator $\Comm$ is invertible
except on states of zero ghost number which are not $\Brst$-exact,
then $\Pcoh{p}$ is trivial except when $p$ is zero.
 \end{Lem}
\Proof Let $V$ denote the subspace of states with non-zero ghost
number. Suppose that $f$ is a state in $V$ such that  $\Brst f=0$
and that $h= \Comm f$ . Then
 \begin{equation}
  f = \Comm^{-1} h = \Comm \Comm^{-2} h.
 \end{equation}
Now, since $\Comm$ is invertible on all states with non-zero ghost
number and exact states of zero ghost number,
 \begin{equation}
  \Brst \Comm^{-2} h = \Comm ^{-1}\Brst \Comm^{-1} h = \Comm^{-1} \Brst f = 0,
 \end{equation}
so that
 \begin{eqnarray}
  f &=&  \Comm \Comm^{-2} h  \End
    &=& \Brst \Gff \Comm^{-2} h.
 \end{eqnarray}
Thus $f$ is cohomologically trivial. \hfil $\Bsq$ \break

The proof of the Fradkin-Vilkovisky theorem is completed by the
observation that (by standard arguments) the path integral
expression (\ref{FVSSeq}) gives the supertrace of the evolution
operator $\Expi{\Hame}$, which by the arguments above must be the
desired generating functional $\trph \Expi{\Hamt}$. The second
non-singularity condition ensures that the sums in the supertrace
may be reordered, in other words the additional term $i\Comm$ in
the Hamiltonian regularises the trace.

\section{A simple example of the theorem}\label{SIMsec}
The first application of the theorem  is to the simple situation
in which the unconstrained phase space is $\Real^4$ (with
coordinates $p_1,p_2,q^1,q^2$) and there is a single first-class
constraint $p_1=0$.  Now it is clear that including the
gauge-fixing condition $q^1=0$ gives the reduced phase space with
coordinates $p_2,q^2$; the constraint simply means that one of the
two dimensions in the configuration space was redundant.  It will
now be seen how this emerges from the \BFV\  formalism in a way
which is useful in more complex situations, following the
``quartet mechanism'' of Henneaux and Teitelboim \cite{HenTei}
which shows that in this case $i\Comm$ is a number operator which
is clearly invertible on states of non-zero ghost number, as well
as on exact states of zero ghost number.

The extended phase space for this system has dimension $(6,4)$ with
local coordinates $p_1,p_2,q^1,q^2, k,l,\Etae,\Thee,\Pie,\Phie$.
Quantization is carried out using a space of states which are
functions of the variables $q^1,q^2,l,\Etae, \Phie$, while
observables are built from the operators $\Op p_1,
\Op p_2, \Op q^1, \Op q^2, \Op k,\Op l, \Op{\Etae},\Op{\Pie} , \Op{\Thee},
\Op{\Phie}$ with
canonical graded commutation relations
 \begin{equation}
  [\Op p_i,\Op q^j] = -i \delta^j_i, \quad [\Op k,\Op l] = -i,
  [\Op{\Etae},\Op{\Pie}] = -i {\rm \ and\ } [\Op{\Thee}, \Op{\Phie}] =i.
 \end{equation}
The operators $\ph_i$ and $\qh^i$ are represented in the standard
way, while the action of $\Op k,\Op l, \Op{\Etae},\Op{\Pie} ,
\Op{\Thee}, \Op{\Phie}$ is defined by
 \begin{eqnarray}
  \Op{k} f(q^1,q^2,l,\Etae, \Phie) &=&- \frac{\partial}{\partial l }
         f(q^1,q^2,l,\Etae, \Phie) \End
  \Op{l} f(q^1,q^2,l,\Etae, \Phie) &=& i l f(q^1,q^2,l,\Etae, \Phie) \End
  \Op{\Etae} f(q^1,q^2,l,\Etae, \Phie) &=& \Etae f(q^1,q^2,l,\Etae, \Phie)\End
  \Op{\Pie} f(q^1,q^2,l,\Etae, \Phie) &=& -i\frac{\partial}{\partial \Etae }
        f(q^1,q^2,l,\Etae, \Phie) \End
  \Op{\Phie} f(q^1,q^2,l,\Etae, \Phie) &=& -i \Phie  f(q^1,q^2,l,\Etae, \Phie) \End
  \Op{\Thee} f(q^1,q^2,l,\Etae, \Phie) &=& \frac{\partial}{\partial \Phie }
       f(q^1,q^2,l,\Etae, \Phie).
 \end{eqnarray}
The inner product is defined by
 \begin{equation}
  (f,g) = \int dq^1\,dq^2\,dl\,d\Etae\, d\Phie\, \Conj{f}(q^1,q^2,-l,\Etae, \Phie)  g(q^1,q^2,l,\Etae, \Phie)
 \end{equation}
where \ $\Conj{}$\  denotes reversal of order of anticommuting
terms together with complex conjugation of complex coefficients so
that $\ph,\qh,\kh, \lh,\Etaeh$ and $\Theeh$ are Hermitian while
$\Phieh$ and $\Pieh$ are antihermitian.  As always in \BRST\
quantization, since the Hermitian operator $\Brst$ has zero
square, an indefinite inner product is required.  The \BRST\
operator and gauge-fixing fermion take the standard forms
 \begin{equation}
  \Brst=\ph_1\Etaeh + \kh\Theeh, \qquad \Gff=\lh\Pieh + \qh^1\Phieh
 \end{equation}
and the commutator is
 \begin{equation}
  \Comm = -i(\ph_1 \lh - \kh \qh^1 + \Etaeh\Phieh + \Theeh\Pieh).
 \end{equation}
Defining creation and annihilation operators
 \begin{eqnarray}
  a&= \Rt{\Op p^1+i \Op k}    \qquad    b&= \Rt{\Op l -i \Op q^1}   \End
   \astar&=\Rt{\Op p^1-i \Op k}  \qquad     \bstar&= \Rt{\Op l + i \Op
 q^1 } \End
  \Cc&=  \Rt{\Op{\Etae} + i \Op{\Thee} } \qquad \Cbar&=\Rt{\Op{\Phie} + i
 \Op{\Pie}} \End
  \cstar&= \Rt{\Op{\Etae} - i \Op{\Thee}}  \qquad \Cbarstar&=\Rt{-
 \Op{\Phie}+i \Op{\Pie}}
\end{eqnarray}
which satisfy commutation relations
 \begin{eqnarray}
  [a,\bstar] &=  [b,\astar] &= 1 \End
  {[\Cc,\Cbarstar]} &=  [\Cbar,\cstar] &= 1
 \end{eqnarray}
(with all other commutators zero), the full space of states is
constructed from  vacuum states $\Vac^u,u=1,2, \dots$ which are
normalized states annihilated by $a,b,\Cc$ and $\Cbar$.
(Explicitly, the vacuum states are
 \begin{equation}
  \Vac^u = \delta\left(\Frac{q-l}{\surd 2}\right)\exp(\Etae\Phie) f_u(q^2)
 \end{equation}
where $\Seto f_u \Setc$ is an orthonormal basis of $L^2(\Real)$.)
The full basis of states
 $\Phin^u$, $u,m,n=1,2, \dots,r,s,=0,1$ is defined by
 \begin{equation}
  \Phin^u =\frac{1}{\sqrt{m!n!}} \astar^m\bstar^n  \cstar^r\Cbarstar^s \Vac^u
 \end{equation}
gives the orthonormality relation
 \begin{equation}
  (\Phin^u,\Phind^{u'}) = \delta_{mn'}\delta_{m'n}\delta_{rs'}\delta_{r's} \delta_{u'u}.
 \end{equation}
The space of states $\Hil$ is then defined to be the space of
states of the form \break
 $\sum_{m,n,r,s,u} a_{mnrsu} \Phin^u$ where
 $\sum_{m,n,r,s,u} |a_{mnrsu}|^2$ is finite.

Observing that the operator $i\Comm$ can be expressed as
 \begin{equation}
  i\Comm= \astar b + \bstar a + \cstar \Cbar + \Cbarstar \Cc
 \end{equation}
which has eigenvalue $m+n+r+s$ on the state $\Phin$, that
 $\cstar \Cbar + \Cbarstar \Cc$
is the operator which gives the sum of the ghost and antighost
number and that all terms in $\Brst$ contain ghost or antighost
creation operators it is clear that the only zero eigenstates of
$\Comm$ have zero ghost number and are not $\Brst$-exact. Hence by
Lemma \ref{COlem} the
\BRST\ charge $\Brst$ has nontrivial cohomology only at ghost
number zero, as required; moreover, provided that the classical
Hamiltonian $\Ham$ is chosen so that $\exp{i\Ham t}$ has a well
defined trace on the reduced phase space, the operator
 $\exp{i(H + i\Comm)t}$
will have a well-defined supertrace, and this will be calculated by
the path integral (\ref{FVSSeq}).

To show how these ideas may be extended to more complex situations
it is useful to introduce the operators $P_{mn}(\kh,\lh), m,n=1,
\dots$ defined inductively by
 \begin{eqnarray}
  P_{00}(\kh,\lh) = 1, \qquad
  P_{m+1\,n}(\kh,\lh) &=&    \kh P_{mn}(\kh,\lh) + P_{mn}(\kh,\lh) \kh, \End
  \quad P_{m\,n+1}(\kh,\lh) &=& \lh P_{mn}(\kh,\lh) + P_{mn}(\kh,\lh) \lh.
 \end{eqnarray}
(It is easily verified that this does consistently define $P_{mn}$
for all non-negative integers $m$ and $n$.) Then, since the vacuum
is annihilated by $a$ and $b$,
 \begin{equation}
  P_{mn}(\kh,\lh) \Vac^u = i^m(\astar)^m (\bstar)^n \Vac^u
 \end{equation}
and also
 \begin{equation}
  i\Comm P_{mn}(\kh,\lh) \Vac^u
  = i(\kh P_{mn}(\kh,\lh) \lh - \lh P_{mn}(\kh,\lh) \kh) \Vac^u
 \end{equation}
so that
 \begin{equation}\label{PMNeq}
   i(\kh P_{mn}(\kh,\lh) \lh - \lh P_{mn}(\kh,\lh) \kh)
   = (m+n+1) P_{mn}(\kh,\lh)
 \end{equation}
as can also be shown algebraically directly from the definition of
$P_{mn}(\kh,\lh)$.

An alternative approach to the space of states for this system,
also developing from the quartet mechanism, is given by Marnelius
and Sandstr\"om \cite{MarSan} in their interesting discussion of
states in \BFV\ quantization, and extended to a wide variety of
systems.
\section{An example to show gauge-fixing in the presence of the Gribov
problem}\label{GRIsec}
An interesting toy example is the system described by Henneaux and
Teitelboim \cite{HenTei} whose initial phase space is $\Real^4$
(with coordinates $p_1,p_2,q^1,q^2$) on which the single first
class constraint
 \begin{equation}
  T \equiv T_1 +T_2 - \textstyle{\frac14} = 0
 \end{equation}
with $T_1 = \Tone$ and $T_2=\Ttwo$, is imposed.
In this case the constraint surface $C$ is the three-dimensional
sphere $S^3$.  The group action on $C$ generated by $T$ is most
easily analysed by defining $z_1=q^1+ip_1$, $z_2=q^2+ip_2$. Then
(with the standard prescription that
 $\delta_{\epsilon}(.)= \epsilon\{T,\ .\}$) $T$ acts as the infinitesimal $U(1)$ transformation
  $\delta z_i = (1+\half i \epsilon z_i)$, giving as true (reduced) phase space of the system
the Hopf fibration $S^2=S^3/U(1)$, which is well known to be
non-trivial  \cite{Nakaha}, so that the system  has a Gribov
problem. This means  that no bosonic gauge-fixing function $X$ can
be found to directly ensure the validity of the Faddeev formula
(\ref{FFeq}). However the  approach developed in this paper does
enable a gauge-fixing procedure to be set up.

To carry out the \BFV\ quantization of this system  the phase space
is extended to
 $S^1 \times \Real^{(5,4)}$ with coordinates $l,k,p_1,p_2,q^1,q^2,\Etae,
\Pie, \Thee$ and $\Phie$, with quantization of these variables as
before, but with a different space of states.

The \BRST\  operator and gauge-fixing fermion are
 \begin{equation}
  \Brst  = T \Etaeh + k\Theeh, \qquad
  \Gff = (\Phieh \Theeh -\Theeh \Phieh) \Pieh \sinh \lh - ( X+ T\cosh \lh)\Phieh
 \end{equation}
where $X= \Xop$.

The space of states is constructed from the states
 $$ P_{mn}(\Th,\kh)\cstar^r\Cbarstar^s \Vac^u, \quad m,n,u = 1,2 \dots, r,s=0,1 $$
where $\Th$ is the operator $\tanh \half \lh$ and the vacuum
states $\Vac^u$ are
 \begin{equation}
  \Vac^u = \exp (\lh T) g_u(q^2) (1+i \Etae \Phie)
 \end{equation}
where $g_u(q^2), u= 1,2 \dots$ are eigenstates of
 $T_2 - \textstyle{\frac14}$ with eigenvalues
$\textstyle{\frac14},\textstyle{\frac34},\dots$ with normalisation
factors set so that the $\Psi_{0,u}$ are orthonormal. The vacuum
states are all annihilated by $\Brst$ and $\Gff$, so that they are
also annihilated by the commutator $\Comm$.

Some algebra shows that (if rows are labeled first by $m-n$ and
then by $m+n$) the  commutator $i\Comm$ is upper triangular with
diagonal elements having  real part $ \sim (m+n+r+s)$. Essentially
the same arguments as in the previous example then shows that the
commutator $\Comm$ does satisfy the nonsingularity conditions.
Then, provided that a suitable Hamiltonian is used, that is, one
which has a finite trace on the space of physical states, the path
integral with Hamiltonian extended by $i\Comm$ will give the
correct supertrace, and thus the desired trace over physical
states.
\section{Conclusion}
This paper establishes a criterion for the admissibility of a
gauge-fixing fermion which can be applied to systems where the
usual criteria are inadequate. In particular it is shown that, by
introducing a gauge-fixing term which is not constructed by the
standard prescription from a bosonic gauge-fixing function, it is
possible to fix the gauge of a system with a Gribov problem.
%


\begin{thebibliography}{10}

\bibitem{fraVil1}
E.S. Fradkin and G.A. Vilkovisky.
\newblock Quantization of relativistic systems with constraints.
\newblock {\em Phys. Lett.}, B55:224, 1975.

\bibitem{BatVil}
I.~A. Batalin and G.A. Vilkovisky.
\newblock Relatavistic {${S}$}-matrix of dynamical systems with boson and
  fermion constraints.
\newblock {\em Phys. Lett.}, B69:309, 1977.

\bibitem{FraFra}
E.S. Fradkin and T.E. Fradkina.
\newblock Quantization of relativistic systems with boson and fermion first-
  and second-class constraints.
\newblock {\em Phys. Lett. B72}, B72:343, 1978.

\bibitem{BatFra}
I.A. Batalin and E.S. Fradkin.
\newblock Operator quantization and abelization of dynamical systems subject to
  first class constraints.
\newblock {\em Revista del Nuovo Cimento}, 9:1--48, 1986.

\bibitem{FraVil2}
E.S. Fradkin and G.A. Vilkovisky.
\newblock Quantization of relativistic systems with constraints: equivalence of
  canonical and covariant formalisms in quantum theory of gravitational field.
\newblock {\em CERN Preprint}, pages TH.2332--CERN, 1977.

\bibitem{Hennea}
M.~Henneaux.
\newblock Hamiltonian form of the path integral for theories with a gauge
  freedom.
\newblock {\em Phys. Rep.}, 126:1, 1985.

\bibitem{HenTei}
M.~Henneaux and C.~Teitelboim.
\newblock {\em Quantization of Gauge Systems}.
\newblock Princeton University Press, 1992.

\bibitem{Govaer1}
J.~Govaerts.
\newblock A note on the {F}radkin-{V}ilkovisky theorem.
\newblock {\em Class. Quant. Grav.}, 8:1723--1746, 1988.

\bibitem{GovTro}
J.~Govaerts and W.~Troost.
\newblock A comparison of {F}addeev and {BFV} phase space path integrals.
\newblock {\em Int. Journ. Mod. Phys.}, A4:4487--4504, 1989.

\bibitem{Schwar}
A.S. Schwarz.
\newblock Lefshetz trace formula and brst.
\newblock {\em Mod. Phys. Lett.}, A4:579, 1989.

\bibitem{McKSin}
H.P. McKean and I.M. Singer.
\newblock Curvature and eigenvalues of the {L}aplacian.
\newblock {\em J. Diff. Geo.}, 1:43--69, 1967.

\bibitem{Gribov}
V.~N. Gribov.
\newblock Quantization of non-{A}belian gauge theories.
\newblock {\em Nuclear Physics}, B130:1--19, 1978.

\bibitem{McMull}
D.~McMullan.
\newblock Constrained quantisation, gauge fixing and the {G}ribov ambiguity.
\newblock {\em Comm. Math. Phys.}, 160:431--456, 1994.

\bibitem{Faddee}
L.D. Faddeev.
\newblock The {F}eynman integral for singular {L}agrangians.
\newblock {\em Teor.i.Mat. Fiz}, 1:3, 1969.
\newblock English translation: {\em Theor. Math. Phys.}, 1:1.

\bibitem{KosSte}
B.~Kostant and S.~Sternberg.
\newblock Symplectic reduction, {BRS} cohomology, and infinite-dimensional
  {C}lifford algebras.
\newblock {\em Annals of Physics}, 176:49--113, 1987.

\bibitem{MarSan}
R.~Marnelius and N.~{Sandstr\"om}.
\newblock Basics of {BRST} quantization on inner product spaces.
\newblock Preprint, October 1998,
\newblock hep-th/9810125.

\bibitem{Nakaha}
M.~Nakahara.
\newblock {\em Geometry, Toplology and Physics}.
\newblock Graduate Student Series in Physics. IOP Publishing, Bristol, UK,
  1990.

\end{thebibliography}
\end{document}